# Hardware Implementation of Keyless Encryption Scheme for Internet of Things Based on Image of Memristors

Mohammad Mohammadinodoushan

Mm3845@nau.edu

Independent Study Report, Advanced Security Laboratory, Northern Arizona University

Spring 2019

**Abstract**— The Internet of Things (IoT) is rapidly increasing the number of connected devices. This causes new concerns towards solutions for authenticating numerous IoT devices. Most of these devices are resource-constrained. Therefore, the use of long-secret keys, in traditional cryptography schemes can be hard to implement. Also, the key generation, distribution, and storage are very complex. Moreover, the goal of many reported cyber-attacks is accessing the key. Therefore, researchers have shown an increased interest in designing keyless encryption schemes recently. In this report, we are going to explain the details of the implementation of the keyless protocol by taking advantage of known technology modules such as microcontrollers (MCU), and hash functions.

Physical Unclonable Functions (PUFs) have been used in many cryptographic applications such as Password Management Systems, key exchange, Key Generation. In this report, we are going to explain the details of the hardware implementation of keyless encryption in the MCU. Different kinds of memristors have been used in the past. In this work, a look-up-table containing memristor cells value at the various current levels is used since the physical component is unavailable yet. The hardware that is used to implement the system is an evaluation-board of SAMV71 MCU, which is used to implement the control system and hardware hashing.

## Introduction

PUFs have been used in many cryptographic applications such as Password Management Systems[1] [2-5] [6], key exchange [7], Key Generation [8], Power Storage [9-14]. In this report, we are going to explain the details of the implementation of keyless encryption based on SHA256 in the microcontrollers and image of memristors. Different kind of memristor exists now, such as MRAMs [15]. Memristor is used as one of the cryptography parts, but we use a look-up-table containing memristor cells value at the different current levels because the physical component is unavailable yet. The hardware used to implement the system is an eval-board of SAMV71 MCU, which is used to implement the control system and hardware hashing. Some steps of the encryption and decryption process are the same and described in the incoming section.

The sender side which encrypts the message and receiver side which decrypts the message use the same RN (Random Number) for each message and share this for every message which is called handshake. Also, both sides use the same password for all the cryptography and share the password in a safe environment at the beginning. Fig. 1 and Fig. 2 show the keyless encryption and decryption protocol, respectively. The code written for simulation and the results of each step is provided in the next sections. The hardware used to implement the system is an eval- board of SAMV71 microcontroller (MCU), which is used to implement the control system and hardware hashing. The evaluation board is shown in Figure 3.

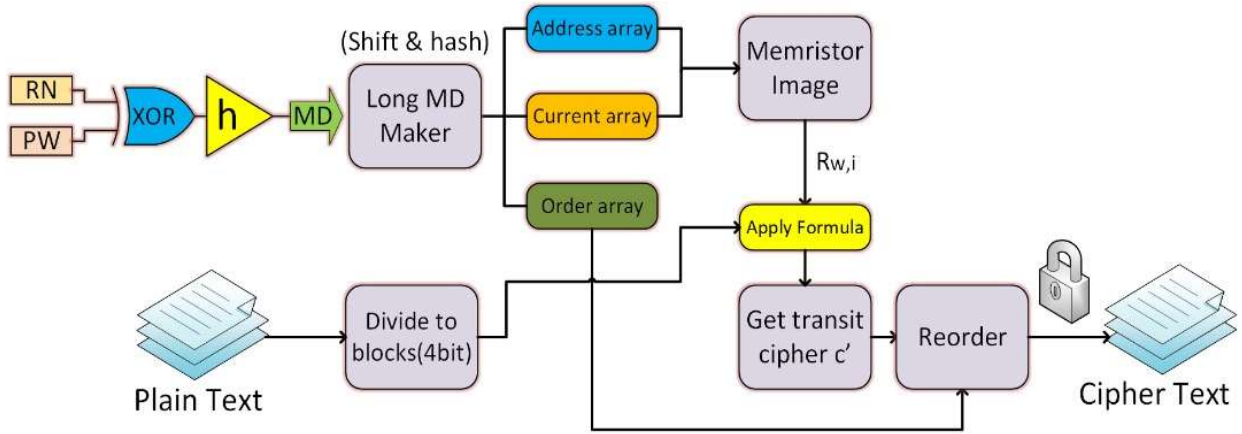

Figure 1: encryption process steps

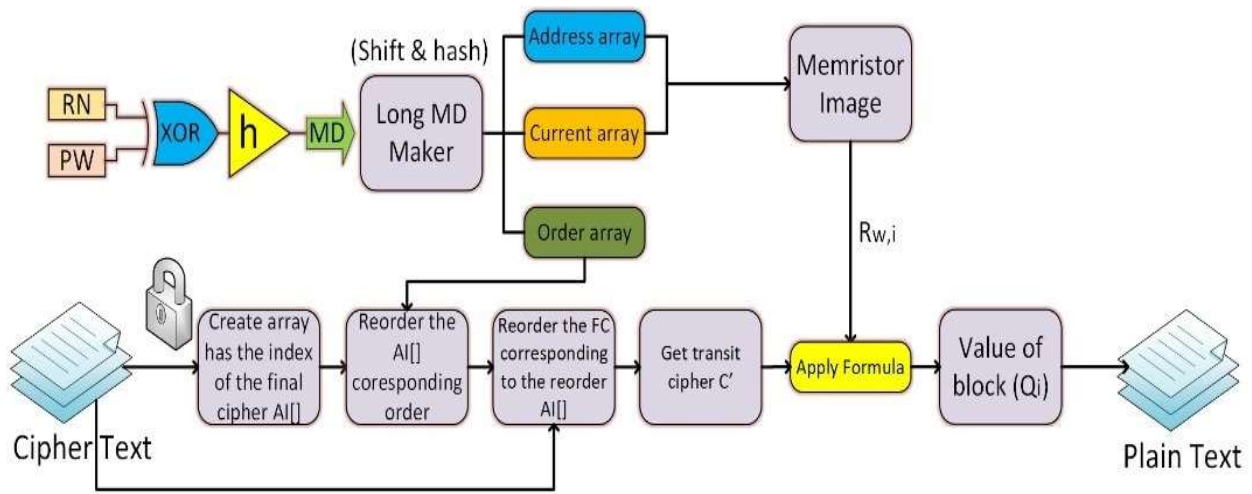

Figure 2: decryption process steps

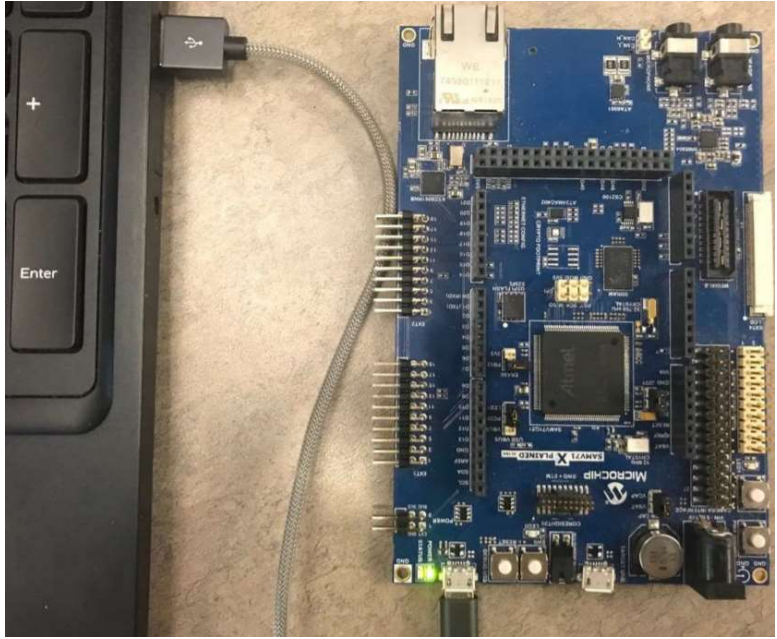

*Figure 3: The evaluation board used to implement the keyless protocol*

## 1.1. Similar steps

Since the keyless encryption and decryption protocol is implemented on the MCU, the similar steps between encryption and decryption are done once at the beginning of the process. XORing, hashing, rotating, creating long MD, and extracting the address, current, and order arrays are similar steps.

### 1.1.1. Generating Long message digest

After handshaking, both parties have the same PW and RN. As it is shown in Fig. 1, at the sender side, Exclusive or (XOR) of ID and PW (ID⊕PW) is calculated. Two modes are considered for the password. The first one is using the saved password, and the second one is entering the password. The second one is just considered to check the results of the code on the MCU with the results of the code developed by software students. Furthermore, three modes are considered for RN: using saved RN, entering a new RN, and generating RN by the hardware engine. Fig. 3. shows the considered cases for a password and a random number.

```
Enter the password:
1- use Saved Pass
2- enter new Pass
1

Enter the Random Number:
1- use saved RN
2- input RN
3- Hardware RN
3

Enter the plain text:
Keyless
```

*Figure 4: Considered cases for password and random number*

In the next step, the hash of (ID⊕PW), h(ID⊕PW), generates the digest. The output of the hash digest is 32 bytes. The left two most significant bytes of the digest is rotated sixteen times, and every time the result is fed into a hash function, and the resulting digest is saved. The longer message digest is 512 bytes, which are the assembly of 16 digests. Rotating, hashing, and generating long MD is done in blow code. Local_var containing two most significant bytes of MD is rotated at each iteration, except at first, and update the sw_digest message and hashing again.

```
for (i_counter = 0; i_counter < 16; i_counter++)
       {
              if(i_counter)
              {
                     local_var = (local_var << 1) | (local_var >> 15);// rotating
                     sw_digest[0] = (local_var >> 8) & 0XFF;
                     sw_digest[1] = (local_var & 0XFF);
              }
              Sha256Calculate(sw_digest, MSG_SIZE, temp_second_hash);
              for(j_count = 0; j_count < MSG_SIZE; j_count++)           // create long MD
              {
                     lmd[i_counter * MSG_SIZE + j_count] = temp_second_hash[j_count];
              }
       }
```

Fig. 4. shows the results of shifting and hashing 16 times. In this figure, the first two bytes are shown in the binary format. This can show more clearly how 16 inputs of hash functions are created by rotating the first two bytes. The results shown in Fig. 4 are obtained after entering "Keyless" as the plain test.

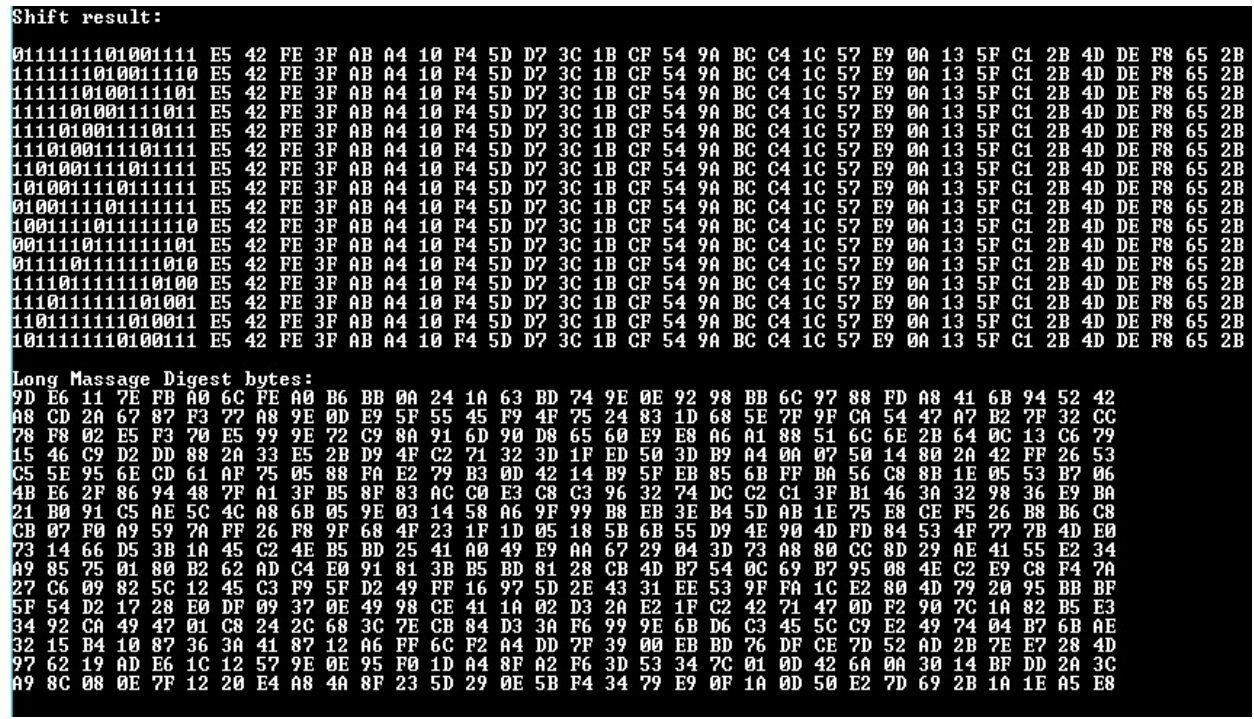

Figure 5: Results of 16 times shifting and hashing

### 1.1.2. Extract Address, current, and order

The long MD is divided into n blocks, each containing address, current, and order. In this implementation, the address and order are 7bit, and the current is 3 bit, and in total, each block is 17 bits. The whole 16*32 bytes of long MD can be used to generate 16*32*8/17 blocks, which is equal to 240.

The address and current are used to access a particular position in the memristor resistance table, which corresponds to the reading of a specific position in the memristor at a certain current level. The order array is used later to reorder to cipher.

## 1.2. Encryption10

The first step in the encryption process is dividing binary representation of the plain text in 4bit blocks, which is called Qi. Then the values read from the Look-up table which is called $R_{w,i}$ and Qi are fed to a block which applies a formula and generates the transit cipher $C'$.

The applied formula is $C' = R_{w,i}(1 + KQ_i)$ and K is equal to 0.2 and constant. The K value is set 0.2 so that the generated ciphers be different enough. The first element of the $C'$ is obtained using $C' = R_{w,i}(1 + 7.5Q_i)$, which is needed for calibration purposes. Fig. 5. Shows the decimal value of the plaintext blocks and transit cipher.

Calculating the first element of transit cipher and the rest of $C'$ elements are calculated in the below code:

```
info_struct[0].transit_cipher = memres_value[0] * (1 + 7.5 * K);

for(i_counter = 1; i_counter < 2*plain_text_size+1; i_counter++)
        {
                info_struct[i_counter].transit_cipher = memres_value[i_counter] * (1 + K * 
                plain_buff[i_counter-1]); // C` = Rw,i*(1+K*Qw,i)
        }
```

```
The string you entered is: Keyless
The Plain text in Hex:4B 65 79 6C 65 73 73 ]
decimal value of each block (4bit) of the plain text: 4 11 6 5 7 9 6 12 6 5 7 3
The transit cipher is: [937 205 492 310 544 350 870 292 295 402 1322 912 257 175
```

*Figure 6: Decimal Value of plaintext blocks and transit cipher*

In the last step, the transit cipher is reordered in corresponding to the order array when the order array is sorting. The output of this reordering step is the final cipher $C$. Fig. 6. shows how the final cipher is extracted from transit cipher using a set of orders.

```
The transit cipher is: [937 205 492 310 544 350 870 292 295 402 1322 912 257 175
The order of cipher before sorting: [121  63  32 125  90  81  38  46   3  76 108
The order of cipher after sorting:  [  3  17  32  33  36  38  46  63  76  81  90
The final cipher is: [295 912 492 257 329 870 292 205 402 350 544 175 1322 937 3
```

*Figure 7: Extracting the final cipher from transit cipher using a set of orders*

In order to manage data well and to deal with arrays during sorting, we packed the address, order, current, and generated transit cipher are packed into structures. Below we use this structure to sort the array and corresponding reordering the transit ciphers:

```
info_struct_t temp_cipher;

for(i_counter = 2*plain_text_size; i_counter ; i_counter--)
{
    for(j = 0; j < i_counter; j++)
    {
        if(info_struct[j].order > info_struct[j+1].order)
        {
```

```
                    temp_cipher = info_struct[j];
                    info_struct[j] = info_struct[j+1];
                    info_struct[j+1] = temp_cipher;
                }
        }
}
```

## 1.3. Decryption

Since a set of orders ($b_{w,i}$) is needed in the decryption, $b_{w,i}$ is saved in a separate array named original_order. The first step in the decryption process is to create a helper index array (HIA). First, the value of HIA is [0,1,2,…,N-2,N-1] where N is the size of the plaintext. In the second step, we sort $b_{w,i}$ array and reorder the HIA corresponding to $b_{w,i}$. Now the orders of HIA and $C$ are changed in the same manner. We sort the HIA and reorder the $C$ to get the transit cipher $C'$.

```
for(i_counter = 2*plain_text_size; i_counter ; i_counter--)
{
        for(j = 0; j < i_counter; j++)
        {
                if(helperindexarray[j] > helperindexarray[j+1])
                {
                        temp_order = helperindexarray[j];
                        helperindexarray[j] = helperindexarray[j+1];
                        helperindexarray[j+1] = temp_order;
                        temp_cipher = info_struct[j];
                        info_struct[j] = info_struct[j+1];
                        info_struct[j+1] = temp_cipher;
                }
        }
}
```

The third step is to apply the inverse of the formula which wasused in encryption process. The invers formula will take the transit cipher $C'$ and output the 4bit blocks of plain text. The Invers formula is $Q_i = (C'/R_{w,i} - 1)/K$ and applies to $C'$ element except the first one which was used for calibration.

The below code shows how we get the 4bit blocks of plain text. One important note in this step that should be considered is that because of dividing floating point number the result may not be real numbers as we expect. We round the output of the inverse formula using roundf function.

```
for(i_counter = 1; i_counter < 2*plain_text_size+1; i_counter++)
{
        local_plain_buff[i_counter-1] = roundf((info_struct[i_counter].transit_cipher /
        memres_value[i_counter] - 1) / K); // Qw,i = (C / Rw,i - 1) / K
}
```

And the final step is concatenating the 4bit blocks to obtain the original plain text.

```
for(i_counter = 0; i_counter < 2*plain_text_size; i_counter+=2)
{
        local_plain_text[i_counter/2] = (local_plain_buff[i_counter] << 4) |
local_plain_buff[i_counter+1];
}
```

The results of this section are shown in Figure 8.

```
The final cipher is: [295 912 492 257 329 870 292 205 402 350 544 175 1322 937 3

Start decrypting process
the helper index array after desort:   [8 11 2 12 14 6 7 1 9 5 4 13 10 0 3 ]
The transit cipher got from final cipher after sorting is: [937 205 492 310 544
4 11 6 5 7 9 6 12 6 5 7 3 7 3
End decryption, the plain text received is: Keyless
```

*Figure 8: Retrieving the plaintext from the cipher*